\documentclass[preprint,showpacs,preprintnumbers,amsmath,amssymb]{revtex4}
\usepackage{graphicx}
\usepackage{dcolumn}
\usepackage{bm}
\usepackage{epsfig}
\usepackage{color}

\begin{document}
\title{Quantum Criticality of  one-dimensional multicomponent Fermi Gas with Strongly Attractive Interaction}

\author{Peng He$^{1}$, Yuzhu Jiang$^{2}$, Xiwen Guan$^{2,3}$, and Jinyu He$^{4}$}

\affiliation{${1}$ Beijing National Laboratory for Condensed Matter
Physics, Institute of  Physics, Chinese Academy of Sciences, Beijing
100190, P. R. China}

\affiliation{${2}$ State Key Laboratory of Magnetic Resonance and
Atomic and Molecular Physics, Wuhan Institute of Physics and
Mathematics, Chinese Academy of Sciences, Wuhan 430071, China}

\affiliation{${3}$ Department of Theoretical Physics, Research
School of Physics and Engineering, Australian National University,
Canberra ACT 0200, Australia}

\affiliation{${4}$ Dezhou University, Dezhou 253023, China}

\begin{abstract}
Quantum criticality of strongly attractive Fermi gas with $SU(3)$
symmetry  in one dimension is studied via the thermodynamic Bethe
ansatz (TBA) equations.
The phase transitions driven by the chemical potential $\mu$,
effective magnetic field $H_1$, $H_2$ (chemical potential biases)
are analyzed at the quantum criticality.
The phase diagram and  critical fields  are analytically  determined
by  the thermodynamic Bethe ansatz equations in zero temperature
limit.
High accurate equations of state, scaling functions are also
obtained analytically for the strong interacting  gases.
The dynamic exponent $z=2$ and correlation length exponent $\nu=1/2$
read off the universal scaling form.
It turns out that the quantum criticality of the three-component
gases involves a sudden change of density of states of one cluster
state,  two or  three cluster states.
In general, this method can be adapted  to deal with the quantum
criticality of multi-component Fermi gases with $SU(N)$ symmetry.

\end{abstract}


\maketitle

\maketitle

\section{Introduction}
Quantum phase transition occurs between different phases of matter
by varying the driving parameter, such as magnetic field, chemical
potential or interaction strength, at zero  temperature.
It is driven by quantum fluctuations associated with the Heisenberg
uncertainty principle rather than by thermodynamic fluctuations
\cite{SachdevBook}.
In critical regime, i.e., the regime near the critical point, the
problem becomes much more difficult because quantum fluctuation and
thermodynamic fluctuation couple strongly with each other.
Novel critical  phenomena associated with rich symmetries  are also
emergent with respect to the quantum phase transition.
For example, one-dimensional(1D) quantum Ising chain with transverse
field exhibits $E_8$ symmetry near the critical point \cite{Coldea}.

However, most of the traditional methods fails in the quantum
critical regime because the fluctuation is very strong and cannot be
neglected.
Therefore, many new methods are proposed in literature,  such as the
effective field theory \cite{SachdevBook, Hertz},  renormalization
group approach \cite{Wison,Senthil},  field theory \cite{Fisher} and
AdS/CFT correspondence \cite{Hartnoll, Herzog, Pires}.
Recently, Zhou and Ho suggested  a practicable  way to map out the
$T=0$ phase diagram of the bulk systems from the density profile of
a trapped gas at low temperatures \cite{Ho-Zhou, ZhouHo}.
This opens the research on quantum criticality in 1D integrable
systems  for which  the equation of state can be systematically
derived in terms of polylogarithm function via thermodynamic Bethe
ansatz (TBA) method in the strong coupling limit \cite{Guan2007,
Erhai, Guan2010pra, hepra}.
The quantum criticality of 1D strongly attractive spin 1/2 Fermi gas
 and the Lieb-Liniger gas were investigated via TBA method \cite{GuanHo,GuanBoson}.
The equation of state obtained analytically from the TBA equations
provides  rich insight into critical behavior.
This method has been applied to the  Bose-Fermi mixture and quantum
gases in an harmonic trap \cite{YinGuan1, YinGuan2}, see review
\cite{GBL}.

Beyond spin-$1/2$ Fermi gas,  the low temperature thermodynamics of
multi-component, especially three-component Fermi gas were studied
for various limited cases
\cite{Schlottmann1,Schlottmann2,Guan2008prl, hepra, Angela}.
The phase diagrams at zero temperature were investigated via both
analytical and numerical methods \cite{Guan2008prl, Angela}.
The finite temperature properties, Tomonaga-Luttinger liquid physics
and equation of state were analytically studied as well
\cite{hepra}.
For three-component Fermi gas, there exist three kinds of composite
particles in attractive regime, which form the normal Fermi gas of
single fermions and quantum gases of  pairs  (two-particle bound
state) and trions (three-particle bound state) as well as their
mixtures.
In contrast, the 1D spin-1 bosons with repulsive density-density and
antiferromagnetic spin-exchange interactions
\cite{Cao,Lee,Shlyapnikov2,Kuhn}  exhibit  either a spin-singlet
paired ground state or a fully polarized ferromagnetic ground state.
These rich phase diagrams provide novel quantum criticality of the
system towards to understanding critical behavior of multicomponent
fermions. The multi-component Fermi gases have been attracted a
considerable attention from a wide range of physics, see a recent
review \cite{Cazalilla2014}.
In particular, recent experiment  with ${}^{171}$Yb atoms with
nuclear spin  $I=1/2$ and ${}^{173}$Yb atoms with $I=5/2$
\cite{Taie} realized the model of Fermi mixture  with $SU(2) \otimes
SU(6)$ symmetry. Realizations of   the  $SU(6)$ Mott-insulator state
with ultracold fermions of ${}^{173}$Yb atoms \cite{Taie:2012} and
the 1D multicomponent fermions of ${}^{173}$Yb \cite{Pagano} open to
further study of ultracold atoms with large spin symmetries.

In this paper, we obtain analytically  higher precision equation of
states which facilitates work out quantum criticality of the model.
The phase boundaries and the scaling functions of density $n$ and
compressibility $\kappa$ are derived in  various  choices of two
external magnetic fields. By controlling two external fields, the
quantum criticality of the model involves different cluster states
of different sizes. This nature can be generalized to multicomponent
Fermi gases.
We further show  that the critical behavior of homogeneous systems
can be mapped out from the density profile of inhomogeneous systems
and the phase boundary at  absolute zero temperature can also be
determined from the finite temperature properties.
In recent years, the experimental simulation with 1D systems develop
fast \cite{Hulet1}.
Our result pave a way to experimental observation of quantum
criticality of  multicomponent Fermi gases.

This paper is organized as follows. In section II, high precision
equation of state is derived in terms of polylogarithm function from
the TBA equations.
 In section III, the phase boundaries of all phases are
given analytically by large-$c$ approximation from the TBA
equations.
In section IV, we studied the quantum criticality of the system at
the phase transitions from vacuum to trion and from the mixture of
unpaired fermions and pairs to the mixture of three kinds of
composite particles, i.e. trions, pairs and single fermions.
In section V, we give a brief summary and discussion.

\section{The Model, TBA equations and equations of states}

The many-body Hamiltonian of a 1D Fermi gas with attractive
$\delta$-function interaction is \cite{Sutherland,Takahashi-2,
Takahashibook}
\begin{equation}
\mathcal{H}_0=-\frac{\hbar ^{2}}{2m}\sum_{i=1}^{N}\frac{\partial
^{2}}{\partial x_{i}^{2}}+g_{1D}\sum_{1\leq i<j\leq N}\delta
(x_{i}-x_{j})+E_z. \label{Ham5}
\end{equation}
where $m$ is the mass of each fermions in this system and the Zeeman
energy is $E_z=\sum_{i=1}^3 N^i \epsilon^i_Z$ here. \cite{hepra} The
contact interaction strength $g_{1D}$ is spin independent and exists
only between fermions with different hyperfine states. It is
negative for attractive interaction  and positive for repulsive
interaction. In the system we considered, there are three possible
hyperfine levels ($\left\vert 1\right\rangle $, $\left\vert
2\right\rangle $, and $\left\vert 3\right\rangle $). The periodic
boundary condition is applied here. For simplicity, we set $\hbar=
2m =1$ and they can be restored  when necessary. In experiments, the
scattering length between the hyperfine states can be tuned via the
broad Feshbach resonance.
 The sublevels can form $SU(3)$ symmetry under a proper choice of  scattering length in each channels.

The Bethe ansatz equations and TBA equations for attractive case are
shown in our earlier work \cite{hepra}. In low temperature, the
contribution of spin fluctuation is suppressed by the strong
magnetic field and can be analytically calculated through our
approach. The full TBA equations read
\begin{eqnarray}
\varepsilon_1&=&k^2-\mu-H_1+T
a_1*\ln(1+e^{-{\varepsilon_2}/{T}})\nonumber\\
&+&T a_2*\ln(1+e^{-{\varepsilon_3}/{T}}) -
T\sum_{n}a_n*\ln(1+\xi_n^{-1}),\label{TBA1}\\
\varepsilon_2& = &2k^2-\frac{c^2}{2}-2\mu-H_2 + T
a_1*\ln(1+e^{-{\varepsilon_1}/{T}})\nonumber\\ &+& T
a_2*\ln(1+e^{-{\varepsilon_2}/{T}}) +T (a_1+a_3) *
\ln(1+e^{-{\varepsilon_3}/{T}})\nonumber\\
 &-& T\sum_{n}a_n*\ln(1+\zeta_n^{-1}), \label{TBA2}\\
\varepsilon_3 & = &3k^2-2{c^2}-3\mu + T
a_2*\ln(1+e^{-{\varepsilon_1}/{T}})\nonumber\\
&+& T(a_1+a_3)*\ln(1+e^{-{\varepsilon_2}/{T}})
+T(a_2+a_4)*\ln(1+e^{-{\varepsilon_3}/{T}}),\label{TBA3}\\
\ln\xi_n&=&\frac{n(2H_1-H_2)}{T}+a_n*\ln(1+e^{-{\varepsilon_1}/{T}})\nonumber\\
&+&\sum_m T_{mn}*\ln(1+\xi_m^{-1})-\sum_m S_{mn}*\ln(1+\zeta_m^{-1}),\label{TBA4}\\
\ln\zeta_n&=&\frac{n(2H_2-H_1)}{T}+a_n*\ln(1+e^{-{\varepsilon_2}/{T}})\nonumber\\
&+&\sum_m T_{mn}*\ln(1+\zeta_m^{-1})-\sum_m
S_{mn}*\ln(1+\xi_m^{-1}), \label{TBA5}
\end{eqnarray}
Here the quantity
$a_{m}\left( x\right) =\frac{1}{2\pi }\frac{m\left\vert c\right\vert
}{\left( mc/2\right) ^{2}+x^{2}}, $ and $\ast$ denotes the
convolution, $(a\ast b)(x)=\int a(x-y)b(y)dy.$ The Eqs.
(\ref{TBA1})-(\ref{TBA3}) are the dressed energies of single atoms
$\varepsilon_1$ and dressed energies of two- and three-body cluster
states $\varepsilon_{2}$ and  $\varepsilon_{3}$ in charge sector.
Here the two-body cluster state involves the two-body bound states
$\{\lambda_j\pm i c/2\}, j=1,\ldots,M_2$ and the three-body bound
states $\{\lambda_j \pm i c, \lambda_j\},j=1,\ldots,M_3$, where
$M_2$ and $M_3$ are the numbers of two-body bound states and
three-body bound states, respectively \cite{hepra}. They are
determined by external fields, chemical potential, interaction
between different clusters and spin wave fluctuations. The
(\ref{TBA4})-(\ref{TBA5}) characterize spin wave fluctuations.
The effective chemical potentials $H_i$ are determined by the
chemical potential $\mu$ and the Zeeman energies \cite{Guan2008prl}.
In very low temperatures, the contributions of spin flipping are
exponential small in strong coupling regimes and therefore they can
be neglected and thus we have
\begin{eqnarray}
\varepsilon _{1}(k) &=& k^{2} -\mu -H_{1} + \, Ta_{1}\ast \ln
(1+e^{-{\varepsilon _{2}(k)}/{T}}) + \, T a_{2}\ast \ln (1+e^{-\varepsilon _{3}(k)/T}), \nonumber \\
\varepsilon _{2}(k) &=& 2k^{2} -2\mu -\frac{c^{2}}{2} -H_{2}+ \,
Ta_{1}\ast \ln (1+e^{-{\varepsilon _{1}}(k)/{T}})
+ \, Ta_{2}\ast \ln (1+e^{-{\varepsilon _{2}}(k) /T}) \nonumber  \\
&&+ \, T(a_{1}+a_{3})\ast \ln (1+e^{-{\varepsilon _{3}}(k)/T}), \nonumber   \\
\varepsilon _{3}(k)  &=&3 k^{2}-3\mu -2{c^{2}}
+Ta_{2}\ast \ln (1+e^{-\varepsilon _{1} (k)/T}) + \, T(a_{1}+a_{3})\ast \ln (1+e^{-{\varepsilon _{2}}(k)/{T}})   \nonumber \\
&&+ \, T(a_{2}+a_{4})\ast \ln (1+e^{-{\varepsilon_{3}}(k)/{T}}).
\label{TBAlowT}
\end{eqnarray}
In the thermodynamic limit, the pressure $p$ is defined as the Gibbs
energy per length \cite{hepra}, which includes three parts, $p_1$,
$p_2$ and $p_3$ and can be expressed in the general form
\begin{equation}
p_r=\frac{r T}{2\pi}\int dk\ln \left(1+e^{-\varepsilon_{r}\left(
k\right)/{T}}\right), \label{pressure}
\end{equation}
where effective masses $r=1,2,3$, which stand for the unpaired
fermions, pairs, and trions, respectively.
Here we have already set the Boltzmann constant $k_{B}=1$.

The TBA equations (\ref{TBAlowT}) are expressed in terms of the
dressed energies $\varepsilon_{1}(k)$, $\varepsilon_{2}(k)$ and
$\varepsilon_{3}(k)$ for unpaired fermions, pairs and trions,
respectively. The dressed energies depend only on the chemical
potential $\mu$ and the external fields $H_1$ and $H_2$ when the
spin terms are neglected in low temperature.
The TBA equations play the central role in the investigation of
thermodynamic properties of exactly solvable models at finite
temperature.  They also provide a convenient formalism to analyze
quantum phase transitions and magnetic effects in the presence of
external fields at zero temperature \cite{review}.

In the strong coupling limit, the convolution integrals in TBA
equations \eqref{TBAlowT} can be simplified and expressed in terms
of the pressure \eqref{pressure} \cite{hepra}
\begin{eqnarray}
\varepsilon _{r}(k) &\approx & r\,k^{2}-A^{(r)},\qquad r=1,\,2,\,3,
\label{epsilon3}
\end{eqnarray}
where $A^{(r)}$ can be written as
\begin{eqnarray}
A^{(1)} &=&\mu +H_{1}-\frac{2}{|c|}p_2 -\frac{2}{3|c|}p_3 +
\frac{1}{4|c|^3}Y_{\frac{5}{2}}^{(2)} +
\frac{1}{9|c|^3}Y_{\frac{5}{2}}^{(3)},
\nonumber \\
A^{(2)} &=&2\mu +\frac12 {c^{2}}+H_{2}
-\frac{4}{|c|}p_1-\frac{1}{|c|}p_2- \frac{16}{9|c|}p_3 +
\frac{8}{|c|^3}Y_{\frac{5}{2}}^{(1)}+
\frac{1}{4|c|^3}Y_{\frac{5}{2}}^{(2)} +
\frac{224}{243|c|^3}Y_{\frac{5}{2}}^{(3)},  \nonumber\\
A^{(3)} &=&3\mu +2c^{2}-\frac{2}{|c|}p_1 -\frac{8}{3|c|}p_2
-\frac{1}{|c|} p_3 + \frac{1}{2|c|^3}Y_{\frac{5}{2}}^{(1)}+
\frac{28}{27|c|^3}Y_{\frac{5}{2}}^{(2)} +
\frac{1}{16|c|^3}Y_{\frac{5}{2}}^{(3)}.
\end{eqnarray}

For simplicity, define
\begin{eqnarray}
Y_a^{(r)}=-\sqrt{\frac{r}{4\pi}}T^a{\rm
Li}_a\left(-e^{A^{(r)}/T}\right),
\end{eqnarray}
where the effective masses $r=1,2,3$. The polylogarithm function is
defined as $\mathrm{Li}_{n}(x)=\sum_{k=1}^\infty \frac{x^k}{k^n}.$
Hence the pressure \eqref{pressure} can be expressed in terms of
polylogarithm function after integration by parts
\begin{eqnarray}
p_1 &=& Y_{\frac{3}{2}}^{(1)} \left[1+\frac{4p_2}{|c|^3} +
\frac{p_3}{3|c|^3}\right] \nonumber \\
&=& Y_{\frac{3}{2}}^{(1)}\left[1+\frac{4}{|c|^3}
Y_{\frac{3}{2}}^{(2)} + \frac{1}{3|c|^3}Y_{\frac{3}{2}}^{(3)}
\right], \nonumber\\
p_2 &=& Y_{\frac{3}{2}}^{(2)} \left[1+ \frac{4p_1}{|c|^3}
+\frac{p_2}{4|c|^3} + \frac{112p_3}{81|c|^3}\right] \nonumber \\
&=&
Y_{\frac{3}{2}}^{(2)}\left[1+\frac{4}{|c|^3}Y_{\frac{3}{2}}^{(1)}
+\frac{1}{4|c|^3}Y_{\frac{3}{2}}^{(2)} + \frac{112}{81|c|^3}
Y_{\frac{3}{2}}^{(3)}
\right], \nonumber\\
p_3 &=& Y_{\frac{3}{2}}^{(3)} \left[1+ \frac{p_1}{3|c|^3}
+\frac{112 p_2}{81|c|^3} + \frac{p_3}{8|c|^3}\right] \nonumber \\
&=&
Y_{\frac{3}{2}}^{(2)}\left[1+\frac{1}{3|c|^3}Y_{\frac{3}{2}}^{(1)}
+\frac{112}{81|c|^3}Y_{\frac{3}{2}}^{(2)} + \frac{1}{8|c|^3}
Y_{\frac{3}{2}}^{(3)} \right].
\end{eqnarray}

Or we can only keep the $1/|c|$ order as
\begin{eqnarray}
p_r=Y_{\frac32}^{(r)}, \qquad r=1,2,3. \label{pressure}
\end{eqnarray}
and
\begin{eqnarray}
p=p_1+p_2+p_3 \label{eqstate}
\end{eqnarray}
is the equation of state. The low temperature thermodynamics can be
studied via this equation in the whole parameter space. One can get
the density $n$ and compressibility $\kappa$ from the equation of
state from the formula $n = \partial p /\partial \mu$ and $\kappa =
\partial^2 p / \partial \mu^2$ directly, which will be shown in the following sections.

\section{The scaling functions of the pure Zeeman splitting case}
The thermodynamics of this system is determined by the TBA
equations, which are usually coupled nonlinear integrated equations.
Thus the approximations are needed in the next step calculations
\cite{hepra}. The properties of 1D Fermi gases in the regime below
crossover temperatures are  described by Tomonaga-Luttinger liquid
theory \cite{Erhai, hepra}. However, in the critical regime, i.e.
near the critical point and above  the crossover temperatures, the
correlation length tends to infinity and the second derivatives of
free energy  becomes divergent. Thus it is difficult to study the
properties near the critical point. In Zhou and Ho's work, the
scaling function read off the critical exponents from
thermodynamical properties at  the quantum criticality. This
provides  a feasible way to study the critical properties in 1D
systems. Guan and Ho applied this method to integrable strongly
attractive Fermi gases with spin $1/2$ to study the quantum
criticality \cite{GuanHo}. We will show that this method can be
applied to the study of multi-component Fermi gases with strongly
attractive interaction.

In the equal Zeeman splitting case, i.e., $H_1=H_2$,  the
three-component problem can be reduced to two-component problem,
i.e., only single fermions and neutral bound states exist, which is
the simplest case of this problem. In former works \cite{hepra}, the
specific heat of the strongly attractive Fermi gases can be
expressed as
\begin{equation}
C_v\sim\left(\frac1{v_u}+\frac{1}{v_t}\right)
\end{equation}
when equal Zeeman field is applied, where $v_u$ and $v_t$ are the
velocity of unpaired fermions and trions, respectively. We can see
that the strongly attractive Fermi gas with equal Zeeman splitting
behaves like two-component Fermi gases with unpaired fermions and
bound trions. This  is a reminiscence of  the properties of spin
half Fermi gas. This suggests us use similar method to deal with the
three-component problems when pure Zeeman splitting exists.

In order to study the quantum criticality, we need to work out the
phase boundaries first. For simplicity, in the following
calculation, we denote the unpaired fermions, pairs and trions as
$A$, $B$, $C$, respectively. For instance, the mixture phase of
trions and pairs are denoted as $B+C$ phase. By analyzing the band
fillings at zero temperature, the phase boundaries  can be obtained
analytically  from the TBA equations. There are three bands in the
three-component problem, corresponding to the unpaired fermions,
bound pairs and trions. For instance, near the boundary between $C$
and $A+C$, the number single fermions is merely equal to zero, while
the population of trions is still large. Thus on the boundary, there
is $\varepsilon_1(0) = 0$ while $\varepsilon_3(Q_3)=0$,  where the
integral boundary $Q_3$ of trion term is given by the Fermi surface,
gives the integral boundary $Q_3$ of trion term, as shown in the
phase boundary equations. Therefore, the phase boundary of this
phase transition is expressed as
\begin{eqnarray}
\mu_{c1} &=&
-H_1-\frac{1}{2\pi}\int_{-Q_3}^{Q_3}\frac{2|c|}{\lambda^2+c^2}\varepsilon_3\left(\lambda\right)d\lambda,\label{mu11}
\end{eqnarray}
where
\begin{eqnarray}
\varepsilon_3(k) &=& 3\left(k^2-\mu-\frac{2 c^2}{3}\right)
-\frac{1}{2\pi}\int_{-Q_3}^{Q_3}\left[\frac{2|c|}{{c^2}
+\left(k-\lambda\right)^2}
+\frac{4|c|}{{4c^2}+\left(k-\lambda\right)^2}\right]
\varepsilon_3\left(\lambda\right)d\lambda, \label{mu12}\\
Q_3^2 &=& \mu+\frac{2 c^2}{3} +\frac{1}{6\pi} \int_{-Q_3}^{Q_3}
\left[\frac{2|c|}{{c^2}+\lambda^2}+\frac{4|c|}{{4c^2}+\lambda^2}\right]
\varepsilon_3\left(\lambda\right)d\lambda.\label{mu13}
\end{eqnarray}
The phase boundaries are determined by these coupled equations.

Similarly, the other three boundaries are
\begin{eqnarray}
\mu_{c2} &=& -\frac{2c^2}{3},\\
\mu_{c3} &=& -H_1,\\
\mu_{c4} &=& -\frac{2c^2}{3} - \frac{2|c|}{3\pi}
\left[\sqrt{\mu_{c4}+H_1}-\frac{1}{|c|}
\left(c^2+\mu_{c4}+H_1\right)\arctan\frac{\sqrt{\mu_{c4}+H_1}}{|c|}\right],
\end{eqnarray}
respectively.

In order to explore the method of calculating scaling functions,
let's consider the simplest case--the phase boundary between the
vacuum and phase $C$. From the equation of state \eqref{eqstate}, we
can  get the density of the system
\begin{eqnarray}
n = 3 Y_{\frac12}^{(3)} \left(1-\frac1{|c|}
Y_{\frac12}^{(3)}\right).\label{equaln1}
\end{eqnarray}
Near the phase boundary, i.e., near the vacuum state, the pressure
$p=p_3$ is very small. Thus it can be neglected during the
calculation and the effective chemical potential $A^{(3)}$  which
can be expressed in terms of $\mu$ and $\mu_c$ as
\begin{eqnarray}
A^{(3)}\approx 3(\mu-\mu_c).
\end{eqnarray}
Here $\mu \approx \mu_c$, and $\mu_c=\mu_{c2}=-2c^2/3$, thus
$A^{(3)}$ is very small. Therefore, the density can be approximately
written in the universal scaling function form \cite{ZhouHo}
\begin{equation}
n(\mu,T) = n_0(\mu,T)+ T^{\frac{d}{z}+1-\frac{1}{\nu z}}\mathcal
{G}\left(\frac{\mu-\mu_c}{T^{\frac{1}{\nu
z}}}\right),\label{scalingfunc}
\end{equation}
where the background value $n_0=0$, the singular function
\begin{equation}
\mathcal {G}= -3\sqrt{\frac{3}{4\pi}}{\rm
Li}_{\frac{1}{2}}(-e^{3(\mu-\mu_{c})/T}),\label{gnb}
\end{equation}
and the critical components are $\frac{1}{\nu z}=1$ and
$\frac{d}{z}+1-\frac{1}{\nu z}=\frac12.$ For 1D systems, the
dimension parameter $d=1$, then it is easy to know that $z=2$ and
$\nu=\frac12$ from the above algebraic equations of critical
components. The following results will show that in all cases the
result of $z$ and $\nu$ are the same, because the critical
components are only determined by the symmetry of the Hamiltonian.
The universal scaling form of Eq. (\ref{scalingfunc}) can be
directly obtained from the equation of states $p=p_1+p_2+p_3$, where
$p_{1,2,3}$ are given by Eq. (\ref{pressure}). However, necessary
approximations are needed in order to obtain the universal scaling
form (\ref{scalingfunc}). Such approximations only involve the
conditions $T\gg \mu - \mu_c$ and $T \ll c/n$ in low temperature
expansions. In contrast, for $T \ll \mu - \mu_c$ and $T \ll c/n$,
the Luttinger liquid thermodynamics is obtained. The explicit
universal scaling forms of other thermodynamical quantities can be
calculated in a similar way, see the compressibility Eq.
(\ref{kappa}) below.

Let's move to another representative case--the phase boundary
between $A+C$ and $C$. In the strong coupling regime, the scaling
function can be obtained under the series expansion and collecting
terms up to $1/|c|$ orders. Near the critical point, we also have
\begin{eqnarray}
A^{(3)} &\approx& 3\left(\mu-\mu_{c}\right)
\end{eqnarray}
by iteration of \eqref{mu12} and \eqref{mu13}, collecting terms up
to order $1/|c|$ and here $\mu_c=\mu_{c1}$.

In quantum critical regime of this phase, there are only a few
trions, thus the pressure $p_3$ is small. From the expression of
pressure, i.e., the equation of state \eqref{eqstate}, we know that
$Y_{\frac12}^{(3)}$ and $Y_{-\frac12}^{(3)}$ are both small.
Meanwhile, the number of pairs is relatively large. Thus by
neglecting some small quantities, the result can be simplified and
the scaling function is obtained as
\begin{eqnarray}
n=n_0 + T^{\frac12} \mathcal{G} \left(\frac{\mu-\mu_c}{T}\right),
\end{eqnarray}
where
\begin{eqnarray}
n_0&=&\frac{4}{\pi}a_{32}^{\frac12}\left(1+\frac{38}{27\pi|c|}a_{32}^{\frac12}\right), \\
\mathcal{G} \left(\frac{\mu-\mu_c}{T}\right)&=&
-3\sqrt{\frac{3}{4\pi}}\left(1 -\frac{32}{9\pi|c|}a_{32}^{\frac12}
\right) {\rm Li}_{\frac12}\left(-e^{{3(\mu-\mu_c)}/{T}}\right).
\end{eqnarray}
Here the approximation
\begin{equation}
{\rm Li}_s\left(-e^u \right) \approx - \frac{u^s}{\Gamma(s+1)}
\end{equation}
has been applied in the above calculation, and
$a_{12}={c^2}/{4}-H_1+{H_2}/{2}$ and $a_{32}=
-{5c^2}/{12}+{H_2}/{2}$. In this  scaling function, $n_0$ is the
contribution from  the background, i.e., the contribution from the
cluster which don't experience a sudden change as the driving
parameter varies across the phase boundaries.

Similarly, the compressibility can be written in the form
\begin{equation}
\kappa - \kappa_0= T^{-\frac12}\mathcal {G}'
\left(\frac{\mu-\mu_c}{T}\right),
\end{equation}
where
\begin{eqnarray}
\kappa_0&=&\frac{2}{\pi}a_{32}^{-\frac12}\left(1+\frac{178}{27\pi|c|}a_{32}^{\frac12}\right), \\
\mathcal{G}' \left(\frac{\mu-\mu_c}{T}\right)&=&
-9\sqrt{\frac{3}{4\pi}}\left(1 -\frac{32}{3\pi|c|}a_{32}^{\frac12}
\right) {\rm Li}_{-\frac12}\left(-e^{{3(\mu-\mu_c)}/{T}}\right).
\end{eqnarray}
Here we can easily see that the critical parameters are still $z=2$
and $\nu=\frac12$.

In general, in the quantum liquid phases of 1D many-body systems,
the equation of states can be written in terms of polylogarithm
functions. Quantum criticality describes universal scaling behavior
of thermodynamics near the critical points. In a small window near a
critical point, the polylogarithm functions capture proper thermal
and quantum fluctuations so that correct critical correct critical
exponents can be mapped out from the scaling forms written in terms
of polylogarithm functions. However, the Sommerfeld expansions with
the equation of states only lead to some terms involving the powers
of temperature, which are not enough to capture such strong thermal
and quantum fluctuations. Therefore, the Luttinger liquid physics
does not contain the critical behavior in the quantum critical
regime.

\section{The general case: The scaling functions at  arbitrary phase boundaries}

In last section, we successfully solve the equal Zeeman splitting
case using  the method for the  spin-half problem. Similarly, we can
also apply this method to solve unequal Zeeman splitting cases even
for the whole parameter plane. We can see that the phase boundary is
determined by $\mu$, $H_1$ and $H_2$. Thus it is possible to give
the general phase boundary equations and the scaling functions.

\subsection{The analytical phase boundaries of different phases in
$\mu-H$ plane}

The above method can also be applied to all the phase boundaries in
this problem. Without loss of generality, we consider the three
states co-exist case--the phase boundary between $A+C$ and $A+B+C$.
Near the phase boundary, the particle number of $B$ is small and
those of $A$ and $C$ are relatively large, hence we have
$\varepsilon_1(Q_1)=0$, $\varepsilon_3(Q_3)=0$, $\varepsilon_2(0)=
0$, where $Q_1$ and $Q_3$ are the integral boundaries of
$\varepsilon_1$ and $\varepsilon_3$, respectively. Thus the phase
boundary is expressed as
\begin{eqnarray}
\mu_c = -\frac{c^2}{4}-\frac{H_2}{2}
-\frac{1}{4\pi}\int_{-Q_1}^{Q_1}\frac{|c|}{\frac{c^2}{4}+\lambda^2}\,
\varepsilon_1  \left(\lambda\right)d\lambda -\frac{1}{4\pi}
\int_{-Q_3}^{Q_3} \left[\frac{|c|}{\frac{c^2}{4}+\lambda^2}
+\frac{3|c|}{\frac{9c^2}{4}+\lambda^2}\right] \varepsilon_3
\left(\lambda\right)d\lambda. \label{ABCB}
\end{eqnarray}
where the $\varepsilon_1$ and $\varepsilon_3$ are determined by the
simplified TBA equations
\begin{eqnarray}
\varepsilon _{1}(k) &=& k^2 -\mu -H_{1}
-\frac{1}{2\pi}\int_{-Q_3}^{Q_3}\frac{2|c|}{{c^2}
+\lambda^2}\,\varepsilon_3 \left(\lambda\right)d\lambda, \label{ABC1} \\
\varepsilon _{3}(k) &=& 3 k^{2}-3\mu -2{c^{2}}
-\frac{1}{2\pi}\int_{-Q_1}^{Q_1}\frac{2|c|}{{c^2}+\lambda^2}\,\varepsilon_1
\left(\lambda\right)d\lambda \nonumber\\
&-&\frac{1}{2\pi}\int_{-Q_3}^{Q_3}\left[\frac{2|c|}
{{c^2}+\lambda^2} +\frac{4|c|}{{4c^2}+\lambda^2}\right]\varepsilon_3
\left(\lambda\right)d\lambda, \label{ABC3}
\end{eqnarray}
respectively, and the integral boundaries $Q_1$ and $Q_3$ satisfy
equations
\begin{eqnarray}
Q_1^2 &=& \mu + H_{1}
+\frac{1}{2\pi}\int_{-Q_3}^{Q_3}\frac{2|c|}{{c^2}
+\lambda^2}\,\varepsilon_3 \left(\lambda\right)d\lambda, \label{ABCQ1} \\
Q_3^2 &=& \mu + \frac{2c^{2}}{3}
+\frac{1}{6\pi}\int_{-Q_1}^{Q_1}\frac{|c|}{{c^2}+\lambda^2}\,\varepsilon_1
\left(\lambda\right)d\lambda
+\frac{1}{6\pi}\int_{-Q_3}^{Q_3}\left[\frac{2|c|}{{c^2}+\lambda^2}+\frac{4|c|}{{4c^2}+\lambda^2}\right]\varepsilon_3
\left(\lambda\right)d\lambda.  \label{ABCQ2}
\end{eqnarray}

The Eqs. \eqref{ABCB}--\eqref{ABCQ2} determine the boundary of phase
transition from $A+C$ to $A+B+C$ together. The last two terms of
\eqref{ABCB} are the effect of the background $A$ and $C$
components, respectively. This is the general method to calculate
all the phase boundaries, and can be applied for such quantum Fermi
gases with $SU(N)$ symmetry when $N$ is an arbitrary integer.

\begin{figure}[t]
\begin{center}
{{\includegraphics [width=0.7\linewidth]{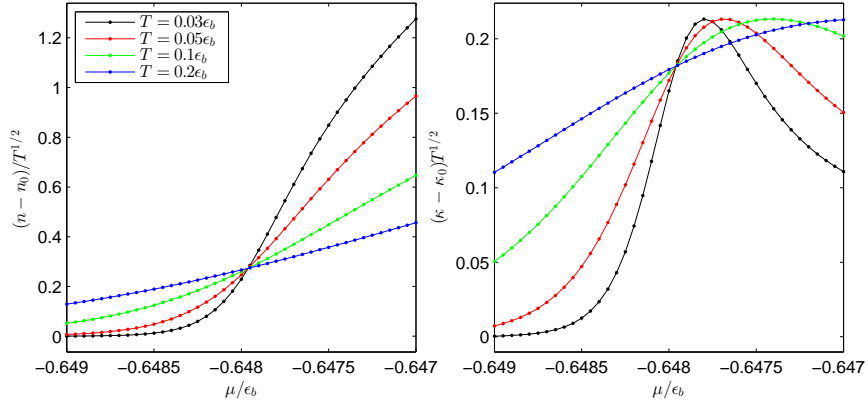}}} \caption{The
left figure shows the  scaled density $n$ vs chemical potential
$\mu$; the right one shows  scaled compressibility $\kappa$ vs
chemical potential $\mu$ for  the phase transition from vacuum to
$B$. In this case, $n_0=\kappa_0=0$. The magnetic fields are set for
equal Zeeman splitting $H_1=H_2=1.32\epsilon_b$. Here we define
$\epsilon_b=c^2/2$. The  curves in each figure are set  for
$T=0.03\epsilon_b, 0.05\epsilon_b, 0.1\epsilon_b, 0.2\epsilon_b$.} \label{figcross1}
\end{center}
\end{figure}
\begin{figure}[t]
\begin{center}
{{\includegraphics [width=0.7\linewidth]{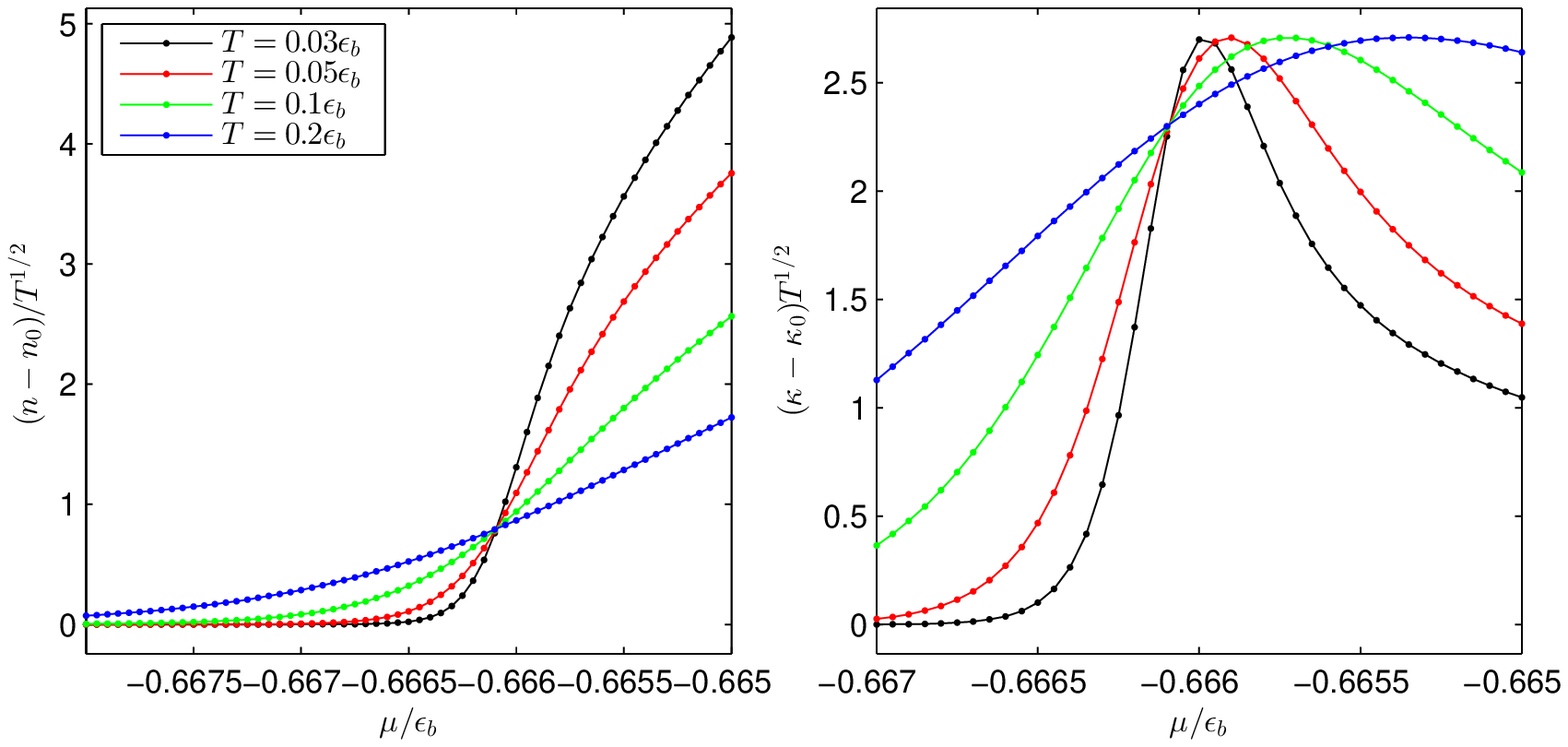}}} \caption{ The
left figure shows the  scaled density $n$ vs chemical potential
$\mu$; the right one shows  scaled compressibility $\kappa$ vs
chemical potential $\mu$ at $T=0.03\epsilon_b, 0.05\epsilon_b, 0.1\epsilon_b, 0.2\epsilon_b$ for  the phase
transition from $B$  to $ B+C$. In this case, $n_0\neq 0,\,
\kappa_0\neq 0$. The magnetic fields are set for equal Zeeman
splitting $H_2=2H_1=1.7\epsilon_b$. } \label{figcross2}
\end{center}
\end{figure}

\begin{figure}[t]
\begin{center}
{{\includegraphics [width=0.7\linewidth]{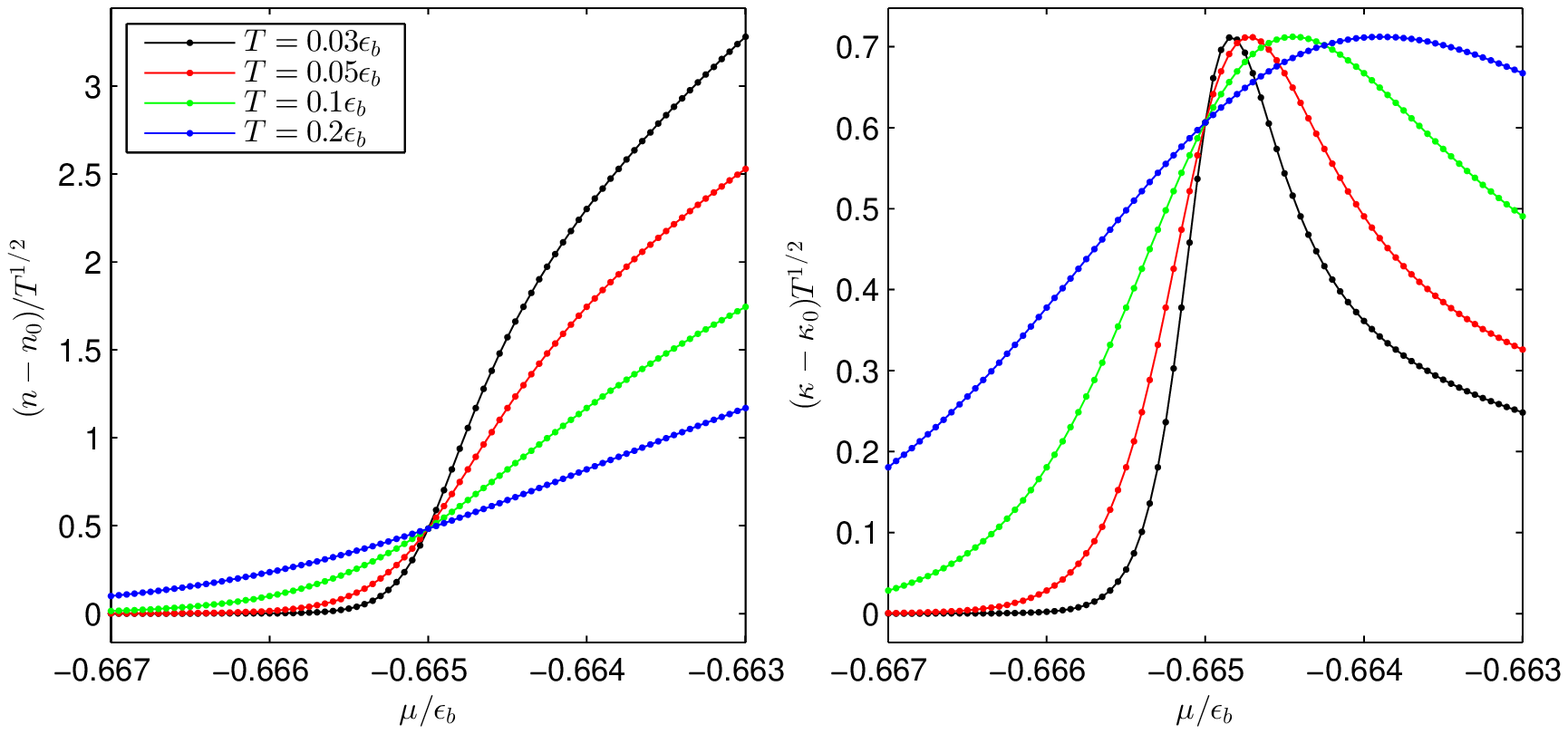}}} \caption{ The
left figure shows the  scaled density $n$ vs chemical potential
$\mu$; the right one shows  scaled compressibility $\kappa$ vs
chemical potential $\mu$ at $T=0.03\epsilon_b, 0.05\epsilon_b, 0.1\epsilon_b, 0.2\epsilon_b$ for  the phase
transition from $A+C $ to $B+C$. In this case, $n_0\neq 0,\,
\kappa_0\neq 0$. The magnetic fields are set for unequal Zeeman
splitting $H_2=1.2H_1$, $H_1=1.34\epsilon_b$. }\label{figcross3}
\end{center}
\end{figure}

\subsection{The universal case of quantum criticality--scaling
functions and phase diagrams}

Now let's calculate the scaling function of the phase transition
from A+C to A+B+C.

The total density and compressibility are
\begin{eqnarray}
n&=&Y_\frac12^{(1)} \left(1-\frac{4}{|c|} Y_\frac12^{(2)}
-\frac{2}{|c|} Y_\frac12^{(3)} \right) +2 Y_\frac12^{(2)}
\left(1-\frac{2}{|c|}Y_\frac12^{(1)}-\frac{1}{|c|}Y_\frac12^{(2)}-\frac{8}{3|c|}Y_\frac12^{(3)}\right)\nonumber\\
&&+3Y_\frac12^{(3)}\left(1-\frac{2}{3|c|}Y_\frac12^{(1)}-\frac{16}{9|c|}Y_\frac12^{(2)}-\frac{1}{|c|}Y_\frac12^{(3)}\right)
\end{eqnarray}
and
\begin{eqnarray}
\kappa&=&Y_{-\frac12}^{(1)}\left(1-\frac{12}{|c|}Y_\frac12^{(2)}-\frac{6}{|c|}Y_\frac12^{(3)}\right)
+4Y_{-\frac12}^{(2)}\left(1-\frac{6}{|c|}Y_\frac12^{(1)}-\frac{3}{|c|}Y_\frac12^{(2)}-\frac{8}{|c|}Y_\frac12^{(3)}\right)\nonumber\\
&&+9Y_{-\frac12}^{(3)}\left(1-\frac{2}{|c|}Y_\frac12^{(1)}-\frac{16}{3|c|}Y_\frac12^{(2)}-\frac{3}{|c|}Y_\frac12^{(3)}\right),
\end{eqnarray}
respectively.

In the limit of $T \rightarrow 0$, $\mu\rightarrow \mu_c$, $n$ and
$T>|\mu-\mu_c|$. Thus $\kappa$ can be cast into  universal form.
Since the $\mu-\mu_c$ terms are much less than the other quantities,
we can keep the zeroth order of $Y_{\frac12}^{(2)}$ and neglect the
first order terms and simplify the $Y_{\frac12}^{(1)}$ and
$Y_{\frac12}^{(3)}$ terms as follows
\begin{eqnarray}
Y_{\frac12}^{(1)} &\approx& \frac{1}{\pi}\sqrt{a_u}, \qquad Y_{-\frac12}^{(1)}  \approx   \frac{1}{2\pi} \sqrt{\frac{1}{a_u}}, \\
Y_{\frac12}^{(3)} & \approx &  \frac{1}{\pi} \sqrt{3 a_t}, \qquad
Y_{-\frac12}^{(3)}  \approx   \frac{1}{2\pi} \sqrt{\frac{3}{a_t}},
\end{eqnarray}
where
\begin{eqnarray}
a_u&=&a_{21}\left(1+\frac{16}{3\pi|c|}a_{21}^{\frac12}
\right)-\frac{32}{9\pi|c|}a_{23}^\frac32, \\
a_t&=&3a_{23}\left(1+\frac{10}{3\pi|c|}a_{23}^{\frac12}
\right)-\frac{8}{3\pi|c|}a_{21}^\frac32,
\end{eqnarray}
and $a_{21}=-{c^2}/{4}+H_1-{H_2}/{2}$ and $a_{23}=
{5c^2}/{12}-{H_2}/{2}$.

Thus we have
\begin{equation}
n - n_0= T^{\frac12}\mathcal {G}
\left(\frac{\mu-\mu_c}{T}\right),\label{n}
\end{equation}
and
\begin{equation}
\kappa - \kappa_0= T^{-\frac12}\mathcal {G}'
\left(\frac{\mu-\mu_c}{T}\right),\label{kappa}
\end{equation}
where
\begin{eqnarray}
n_0 &=& \frac{\sqrt{a_u}}{\pi} \left(1
-\frac{2\sqrt{3}}{\pi|c|}\sqrt{a_t} \right) +
\frac{3\sqrt{3}}{\pi}\sqrt{a_t} \left(1 - \frac{2}{3\pi|c|}
\sqrt{a_u} -\frac{\sqrt{3}}{\pi|c|} \sqrt{a_t} \right),\label{n0-abc} \\
\mathcal {G} \left(\frac{\mu-\mu_c}{T}\right) &=&
-\sqrt{\frac{2}{\pi}}\left(1 - \frac{2}{\pi|c|} \sqrt{a_u}
-\frac{8\sqrt{3}}{3\pi|c|} \sqrt{a_t} \right) {\rm
Li}_{\frac12}\left(-e^{{2(\mu-\mu_c)}/{T}} \right),
\end{eqnarray}
and
\begin{eqnarray}
\kappa_0 &=& \frac{1}{2\pi\sqrt{a_u}} \left(1
-\frac{6\sqrt{3}}{\pi|c|}\sqrt{a_t} \right) +
\frac{9\sqrt{3}}{2\pi\sqrt{a_t}} \left(1 - \frac{2}{\pi|c|}
\sqrt{a_u} -\frac{3\sqrt{3}}{\pi|c|} \sqrt{a_t} \right), \\
\mathcal {G}' \left(\frac{\mu-\mu_c}{T}\right) &=&
-2\sqrt{\frac{2}{\pi}}\left(1 - \frac{6}{\pi|c|} \sqrt{a_u}
-\frac{8\sqrt{3}}{\pi|c|} \sqrt{a_t} \right) {\rm
Li}_{-\frac12}\left(-e^{{2(\mu-\mu_c)}/{T}} \right).\label{g-abc}
\end{eqnarray}

\begin{figure}[t]
\begin{center}
{{\includegraphics [width=0.7\linewidth]{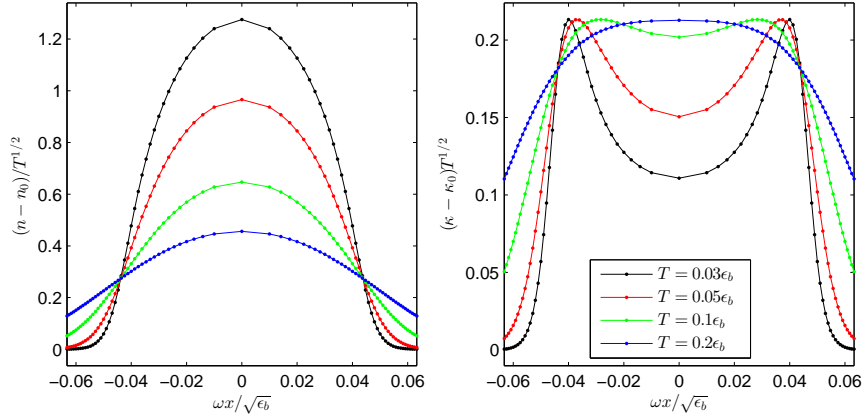}}}
\caption{\label{fig:harm} Quantum criticality can be mapped out by
the measurement of local values of thermodynamics  in a harmonic
trap. The left panel  shows the local density $n(x)$ vs
dimensionless position $\omega x$ for the phase transition from
vacuum into the phase $B$ at $T=0.03\epsilon_b, 0.05\epsilon_b,
0.1\epsilon_b, 0.2\epsilon_b$. Whereas the right one shows the local
compressibility $\kappa(x)$ vs $\omega x$  at the same temperatures
$T=0.03\epsilon_b, 0.05\epsilon_b, 0.1\epsilon_b, 0.2\epsilon_b$.
Here numerical settings are $\mu_0=-0.647\epsilon_b$ and
$H_1=H_2=1.32\epsilon_b$. For this phase transition,  the regular
parts of the density and compressibility are both zero. From proper
temperature-scaled density and compressibility, one reads off the
critical exponents $Z=2$ and $\nu=1/2$.  With such exponents, the
density and compressibility indeed show  the intersection nature at
the critical point. }
\end{center}
\end{figure}
Compare with the first case, we can see the difference in   the
background: state $B$ is added from vacuum phase in the first case
and in this case it is added from $A+C$ phase. The first and second
terms of $n_0$ and $\kappa_0$ are the contributions from unpaired
fermions and trions, respectively. The $a_u$ and $a_t$ terms in
$\mathcal {G}$ and $\mathcal {G}'$ are  the contributions from $A$
and $C$, respectively. Again we can see that the critical parameters
are still $z=2$ and $\nu=\frac12$. From the scaling functions of $n$
and $\kappa$ we have calculated above, we can plot the scaled
diagrams of $\left[{(n-n_0)}/\sqrt{T}\right]-\mu$ and
$\left[{(\kappa-\kappa_0)}\sqrt{T}\right]-\mu$, see Fig.
\ref{figcross1}-Fig. \ref{figcross3}. The diagrams show that the
curves intersect at the critical  point. The cross point gives one
point on the boundary between two different phases. This method  map
out the phase diagrams of homogeneous systems through the trapped
ultracold atoms in experiments.

In the trapped gas, the local chemical potential $\mu$ is replaced
by $\mu(x)=\mu_0-\frac12 m \omega^2 x^2$. Within the local density
approximation, the quantum criticality can be mapped out from the
trapped gas at finite temperatures. For example, for the phase
transitions from vacuum to $B$, the quantum criticality of density
and compressibility can be mapped out by the measurement of the the
local values, see Fig \ref{fig:harm}. Similar study can be carried
out for other critical regions.

\begin{figure}[t]
\begin{center}
\includegraphics [width=0.7\linewidth]{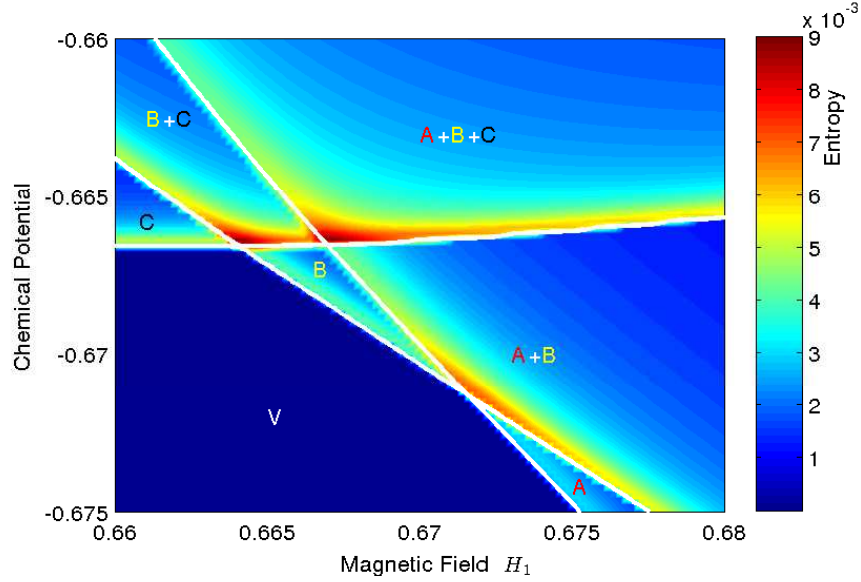}
\caption{The phase diagram of the entropy $S$ in $\mu-H_1$ plane for
temperature   $T=0.001\epsilon_b$. Here $H_2=1.255 H_1$. The
white lines show the phase boundaries at $T=0$, see Section III.
}\label{fig:muH}
\end{center}
\end{figure}

The phase diagrams at finite temperatures can also be achieved from
the equation of state, see Fig. \ref{fig:muH}. From this figure we
see that one can control two external fields $H_{1}$ and $H_2$
trigger multiple critical points. So that one can have phase
transition from one cluster state into multiple cluster states, for
example from phase $B$ into the phase $A+B+C$, as the chemical
potential increase across the multiple critical point.   Here the
scaling function of thermodynamics involves the background of B
states and the singular part consisting of three cluster states.
This method is also universal and can be applied to Fermi gases with
arbitrary $SU(N)$ symmetry.

\section{Conclusion}
In conclusion, polylogarithm functions have been  applied to study
the quantum critical behavior of 1D strongly attractive
three-component Fermi gas with both linear and nonlinear Zeeman
splitting at  low temperatures via thermodynamic Bethe ansatz
equations (\ref{TBA1}) to (\ref{TBA5}).
The equation of state, phase boundaries and scaling functions have
been  derived analytically  in terms of  interaction strength
 $1/|c|$.
From the scaling functions, the phase diagrams at zero temperature
for homogeneous systems can be mapped out from the inhomogeneous
trapping systems at finite temperatures.
The general forms of quantum criticality (\ref{n}) and (\ref{kappa})
can be written  in terms of  the multiple  changes of cluster states
either in regular part or in singular part.
For example the density and compressibility can be cast into the
following general forms
\begin{eqnarray}
&n - n_0= T^{\frac12}\mathcal {G}
\left(\frac{\mu-\mu_c}{T}\right),~~~ \mathcal {G}(x)=-\sum_{j=1}^N
\frac{r_j^{3/2}}{2\sqrt{\pi}}\lambda_j {\rm Li}_{1/2}(-{\rm
e}^{r_jx}),
\label{n}\\
&\kappa - \kappa_0= T^{-\frac12}\mathcal {G}'
\left(\frac{\mu-\mu_c}{T}\right), \mathcal {G}'(x)=-\sum_{j=1}^N
\frac{r_j^{5/2}}{2\sqrt{\pi}} \lambda'_j {\rm Li}_{-1/2}(-{\rm
e}^{r_jx}), \label{kappa}
\end{eqnarray}
where the background density $n_0$,  compressibility $\kappa_0$
involve the constant parts of these quantities which only depend on
the critical effective magnetic fields and chemical potentials,
$r_j$ are the particle effective masses of the corresponding
classters with the sudden change of the density of the states,
$\lambda_j$ are some constants and the addition is taken over all
the driven strings. When the background of the phase transition is
the vacuum state, $\lambda_j=1$.

Whereas the scaling functions of the singular parts $\mathcal{G}$
and $\mathcal{G}'$ are uniquely determined by a sudden change of
density  of state of  the single and/or cluster states.
For example, Eqs. (\ref{n0-abc})-(\ref{g-abc}) near the phase
transition from $A+C$ to $A+B+C$ illustrate the case involving the
sudden change of the density of state of the two-atom cluster
states.
However, for the phase transition from phase $B$ to $A+B+C$, the
scaling function involves two parts related to  the sudden change of
density  of state of  the single fermions and three-body  cluster
states, see Fig.~\ref{fig:muH}.
The phase diagram at the  zero temperature can be mapped out from
the trapped gas at finite temperatures with  quantum criticality.
This method is in general valid for 1D  continuous Fermi gases  and
can be applied to study the universal quantum criticality of
strongly attractive Fermi gases with high spin symmetries.
As an obvious example, the general $SU(N)$ case can be well studied
via this method.
Moreover, the far-from equilibrium universal dynamics of one
dimensional interacting fermions with large spin symmetries is
particularly interesting. One can expect  that the non-equilibrium
states of the large spin systems crossing of a phase transition are
described by the Kibble-Zurek mechanism, see  a recent  review
\cite{De-Campo}.  Integrable models out of equilibrium crossing
quantum phase transitions  would provide practicable settings for
the  Kibble-Zurek mechanism.

\acknowledgments

This work is in part supported by the National Basic Research
Program of China under Grant number 2012CB922101 and  NSFC under
Grant number  11374331 and 11304357,   the Knowledge Innovation
Project of Chinese Academy of Sciences and the Australian Research
Council. The authors  thank Prof. Yupeng Wang for helpful
discussion.

\end{document}